\documentstyle[12pt,aaspp4]{article}
\lefthead{Abramenko et al.}
\righthead{Turbulent Diffusion in the Photosphere}

\begin{document}

\title{Turbulent Diffusion in the Photosphere as Derived from Photospheric
Bright Point Motion}

\author{Abramenko, V.I. $^1$, Carbone, V.$^2$, Yurchyshyn, V.$^1$, Goode,
P.R.$^1$, Stein, R.F.$^3$, Lepreti, F.$^2$, Capparelli, V.$^2$, Vecchio, A.$^2$
}

\affil{
$^1$ Big Bear Solar Observatory, 40386 N. Shore Lane, Big Bear City, CA
92314, USA;\\
$^2$ Dipartimento di Fisica, Universita della Calabria-
Via P. Bucci 31/C, I-87036 Rende, Italy;\\
$^3$ Department of Physics \& Astronomy, Michigan State University,
 East Lansing, MI 48824, USA}

\begin{abstract}
On the basis of observations of solar granulation obtained with the New Solar
Telescope (NST) of Big Bear Solar Observatory, we explored proper motion of
bright points (BPs) in a quiet sun area, a coronal hole, and an active region
plage. We automatically detected and traced bright points (BPs) and derived
their mean-squared displacements as a function of time (starting from the
appearance of each BP) for all available time intervals. In all three magnetic
environments, we found the presence of a super-diffusion regime, which is the
most pronounced inside the time interval of 10-300 seconds. Super-diffusion,
measured via the spectral index, $\gamma$, which is the slope of the
mean-squared displacement spectrum, increases from the plage area
($\gamma=1.48$) to the quiet sun area ($\gamma=1.53$) to the coronal hole
($\gamma=1.67$). We also found that the coefficient of turbulent
diffusion changes in direct proportion to both temporal and spatial scales. For
the minimum spatial scale (22 km) and minimum time scale (10 sec), it is 22 and
19 km$^{2}$ s$^{-1}$ for the coronal hole and the quiet sun area, respectively,
whereas for the plage area it is about 12 km$^{2}$ s$^{-1}$ for the minimum time
scale of 15 seconds. We applied our BP tracking code to 3D MHD model data of
solar convection (Stein et al. 2007) and found the super-diffusion with
$\gamma=1.45$. An expression for the turbulent diffusion coefficient
as a function of scales and $\gamma$ is obtained.
\end{abstract}


\section{Introduction}

Diffusion of magnetic elements on the solar surface reflects
sub-photospheric turbulent convection. It is an energy source for heating upper
layers of the solar atmosphere and is essential for dynamo action. These
circumstances justify the persistent interest of scientists to various problems
related to turbulent diffusion of solar magnetic fields.

Recent observations of the solar magnetic surface with high powered space- and
ground-based instruments demonstrated, beyond doubt, that the quiet sun (QS)
magnetic fields can no longer be considered to be sheets of unipolar magnetic
flux stretching along the boundaries of supergranules. Hinode instruments
(Kosugi et al. 2007; Tsuneta et al. 2008) facilitated a significant breakthrough
in our understanding of QS magnetism: Ubiquitous transverse magnetic fields
(e.g., Lites et al. 2008; Ishikawa et al. 2010; Ishikawa \& Tsuneta 2010) and
fine intranetwork mixed polarity fields (e.g., de Wijn et al. 2008, 2009) were
discovered. The launch of the IMaX spectropolarimeter on-board the Sunrise
balloon-borne solar observatory (Martinez Pillet et al. 2011) allowed one the
resolution of individual kG flux (Lagg et al. 2010) and exploration of vortex
flow motions (Bonet et al. 2010), to mention a few (for more results from
IMaX/Sunrise experiment see a special issue of The Astrophysical Journal
Letters, 713).

All of this new information suggests that the QS magnetic fields can not be
fully explained as remnants of decaying active regions but rather leads to the
idea that an additional mechanism for magnetic field generation, such as local
turbulent dynamos (see review by Petrovay 2001, Pietarila Graham et al. 2010,
and references herein), should be at work. One of the essential parameters for
generating magnetic fields via turbulent dynamo action is the magnitude of
magnetic diffusivity: in the case of very high diffusivity, chances for magnetic
field concentrations to resist the spreading action of turbulent flows are small
and, so, the dynamo could be restrained.

Another important question is how diffusivity varies with spatial and time
scales?
The commonly accepted mechanism that transports magnetic flux over the
solar surface is random walk, or, \textit{ normal diffusion}, when the
mean-squared displacements of flow tracers vary with time, $\tau$, as $\langle
(\Delta l)^2 \rangle = 4K\tau \sim \tau^{\gamma}$, where $K$ is the diffusion
coefficient (a scalar), and $\gamma=1$. In this case diffusivity does not vary
with spatial and temporal scales. Generally, when index $\gamma$ deviates from
unity, the diffusion is then called \textit{ anomalous}. More specifically, the
regime with $\gamma>1$ is called \textit{ super-diffusion} (the diffusion
coefficient growths with scales), while $\gamma<1$ indicates \textit{
sub-diffusion} (the diffusion coefficient decreases when the scales increase,
see Section 6).

From tracking magnetic elements, Lawrence \& Schrijver (1993) reported an
anomalous sub-diffusion on supergranular spatial scales and time scales of days.
They found the diffusivity spectral index $\gamma=0.89 \pm 0.20$, which is
smaller than unity and thus implies sub-diffusion. Cadavid et al. (1999) tracked
G-band bright points and found sub-diffusivity ($\gamma=0.76 \pm 0.04$) on
granular scales and time intervals of 0.3-22 minutes and nearly normal diffusion
($\gamma=1.10 \pm 0.24$) for 25-57 minutes. Later, Lawrence et al. (2001),
utilized the same G-band data set by applying subsonic filtering (i.e., removing
displacements with velocities larger than 7 km s$^{-1}$ - the photospheric sound
speed), and they reported super-diffusion with $\gamma=1.27 \pm 0.01$ on time
scales of 0.8-34 minutes. They also found that in the frameworks of the
continuous time random walk approach (CTRW, see, e.g., Campos \& Mendez 2008;
Zoia et al. 2010, see also references in Cadavid et al. 1999), the diffusion
exponent is $1.54 \pm 0.39$

Above we discussed studies that explored the regime of anomalous diffusion only,
whereas a voluminous body of literature exists that deals with derivation of
diffusivity coefficient in the framework of normal diffusion.

For example, Hagenaar et al. (1999) analyzed
displacements of magnetic elements and found that the magnetic diffusion
coefficient also depend on the time scale: it is about 70-90 km$^{2}$s$^{-1}$
when measured over time intervals of 1-3 hours and it increases to 200-250
km$^{2}$s$^{-1}$ for time intervals longer than 8 hours. Berger et al. (1998b)
tracked G-band BPs observed with the Swedish Vacuum Solar Telescope (SVST) and
found a decreasing diffusion coefficient for time scales growing from 23 sec to
approximately 25 min, while there was a slight increase for longer time scales
(from 50 to 79 km$^{2}$s$^{-1}$ on time scales from 27 to 57 minutes). 

Chae et al. (2008) reported an increase in the diffusion coefficient as the
smallest resolved spatial scale increases. A review of the photospheric magnetic
diffusion covering the pre-Hinode era is given in Schrijver et al. (1996), 
where the authors emphasize a significant discrepancy between the previously
reported diffusivity values. The reasons for that seem to be  an acute
sensitivity of the derived values to the temporal and spatial resolution of
data, as well as the intrinsic dependence of the rate of turbulent diffusion on
the spatial scale (see, e.g., Monin \& Yaglom 1975; Chae et al. 2008).

In the present study, we focus on properties of magnetic diffusion in the quiet
sun photosphere, especially on their changes with varying temporal and spatial
scales. Bright features, visible in the dark inter-granule lanes, are called
bright points (BPs) and they are thought to be footpoints of magnetic flux tubes
(e.g., Muller et al. 2000; Berger \& Title 2001; Ishikawa et al. 2007, to
mention a few). Therefore, they make it possible for us to measure the dynamics of
the photospheric magnetic flux tubes. Only a fraction (about 20\%, de Wijn et
al. 2008) of magnetic elements are thought to be associated with BPs, therefore
they allow us to study only a subset of the entire magnetic flux tube
population. For that purpose, we will use the highest resolution data on
solar granulation (see Table 1) obtained with the New Solar Telescope (NST,
Goode et al. 2010a) of the Big Bear Solar Observatory (BBSO).

\begin{table}[!ht]
 \caption{\textsf{Parameters of Solar Instruments}}

\begin{centering}
\textsf{\footnotesize }\begin{tabular}{lccccclll}
\hline
\textsf{\footnotesize Name } & \textsf{\footnotesize D, cm } & \textsf{\footnotesize $\lambda$, nm } & \textsf{\footnotesize Diff limit } & \textsf{\footnotesize Pixel } & \textsf{\footnotesize Cadence, s } & \textsf{\footnotesize AO$^a$ } & \textsf{\footnotesize Speckle} & \textsf{\footnotesize Reference }\tabularnewline
\hline
\textsf{\footnotesize SOHO/MDI/HR } & \textsf{\footnotesize 10 } & \textsf{\footnotesize 676.7 } & \textsf{\footnotesize 1.$''$70 } & \textsf{\footnotesize 0.$''$6 } & \textsf{\footnotesize 60 } & \textsf{\footnotesize NO } & \textsf{\footnotesize NO } & \textsf{\footnotesize Sherrer et al. 1995 }\tabularnewline
\textsf{\footnotesize SDO/HMI } & \textsf{\footnotesize 14 } & \textsf{\footnotesize 617.3 } & \textsf{\footnotesize 1.$''$11 } & \textsf{\footnotesize 0.$''$5 } & \textsf{\footnotesize 45 } & \textsf{\footnotesize NO } & \textsf{\footnotesize NO } & \textsf{\footnotesize http://hmi.stanford.edu/ }\tabularnewline
\textsf{\footnotesize Hinode/SOT } & \textsf{\footnotesize 50 } & \textsf{\footnotesize 430.5 } & \textsf{\footnotesize 0.$''$216 } & \textsf{\footnotesize 0.$''$109 } & \textsf{\footnotesize 60 } & \textsf{\footnotesize NO } & \textsf{\footnotesize NO } & \textsf{\footnotesize Tsuneta et al. 2008 }\tabularnewline
 &  & \textsf{\footnotesize 630.2 } & \textsf{\footnotesize 0.$''$316 } & \textsf{\footnotesize 0.$''$16 } & \textsf{\footnotesize 120 } & \textsf{\footnotesize NO } & \textsf{\footnotesize NO } & \textsf{\footnotesize Tsuneta et al. 2008 }\tabularnewline
\textsf{\footnotesize SVST } & \textsf{\footnotesize 50 } & \textsf{\footnotesize 430.5 } & \textsf{\footnotesize 0.$''$216 } & \textsf{\footnotesize 0.$''$083 } & \textsf{\footnotesize 23 } & \textsf{\footnotesize NO } & \textsf{\footnotesize NO } & \textsf{\footnotesize Berger et al. 1998b }\tabularnewline
\textsf{\footnotesize SST } & \textsf{\footnotesize 100 } & \textsf{\footnotesize 430.5 } & \textsf{\footnotesize 0.$''$108 } & \textsf{\footnotesize 0.$''$041} & \textsf{\footnotesize 20 } & \textsf{\footnotesize YES } & \textsf{\footnotesize YES } & \textsf{\footnotesize Scharmer et al. 2003 }\tabularnewline
\textsf{\footnotesize DST/ROSA } & \textsf{\footnotesize 76 } & \textsf{\footnotesize 430.5 } & \textsf{\footnotesize 0.$''$142 } & \textsf{\footnotesize 0.$''$069 } & \textsf{\footnotesize 23 } & \textsf{\footnotesize YES } & \textsf{\footnotesize YES } & \textsf{\footnotesize Crockett et al. 2010 }\tabularnewline
\textsf{\footnotesize IMaX/SUNRISE } & \textsf{\footnotesize 100 } & \textsf{\footnotesize 525.0 } & \textsf{\footnotesize 0.$''$132 } & \textsf{\footnotesize 0.$''$055 } & \textsf{\footnotesize 33 } & \textsf{\footnotesize NO } & \textsf{\footnotesize NO } & \textsf{\footnotesize Barthol et al. 2011 }\tabularnewline
\textsf{\footnotesize BBSO/NST } & \textsf{\footnotesize 160 } & \textsf{\footnotesize 705.7 } & \textsf{\footnotesize 0.$''$110 } & \textsf{\footnotesize 0.$''$0375 } & \textsf{\footnotesize 10 } & \textsf{\footnotesize YES } & \textsf{\footnotesize YES } & \textsf{\footnotesize Goode et al. 2010b }\tabularnewline
\hline
\end{tabular}
\end{centering}
\footnotesize{$^a - $ adaptive optics correction.}
\end{table}

\section{Data and Analysis}

In this study, we analyzed three different data sets: i) a quiet sun
internetwork/network area (QS), ii) a coronal hole area (CH), and iii) a plage
area inside an active region (ARP).

Solar granulation was observed with the NST, which is a 1.6 meter clear aperture
off-axis reflector (Goode et al. 2010a,b). The observations were taken with a
1~nm bandpass TiO interference filter centered at a wavelength of 705.7 nm.
Previously, photospheric BPs were studied with G-band data (e.g., Berger et al.
1998a,b; Cadavid et al. 1999; Utz et al. 2009, 2010; Crockett et al. 2010;
Sanchez Almeida et al. 2010, to mention a few) obtained using an interference
filter centered at 430.5 nm. Direct comparison of simultaneous Hinode G-band and
NST/TiO images of the same area (see Figure 1 in Abramenko et al. 2010) showed
that although there is a difference in BP intensities, these two spectral ranges
are equally suitable for detecting photospheric BPs.

The QS data set was obtained during an uninterrupted observing run between 17:06
and 18:57 UT on August 3, 2010, when the telescope was pointed at the disk
center and seeing conditions were excellent. A total of 648 images with a 10 second
cadence were obtained. The CH data set consists of 183 images taken with a 15
second cadence obtained on August 31, 2010 between 17:41 UT and 18:15 UT. The
telescope was pointed at a confined CH located at E12N02. The ARP data set
has 513 images with a 10 second cadence of AR NOAA 11109 recorded on
September 27, 2010 between 17:22 UT and 19:51 UT. The CH and ARP sets were both
recorded at very good seeing conditions. All data sets were adaptive optics (AO)
corrected, speckle reconstructed and destreched.

\begin{figure}[!h]
\centerline{\epsfxsize=7.0truein\epsffile{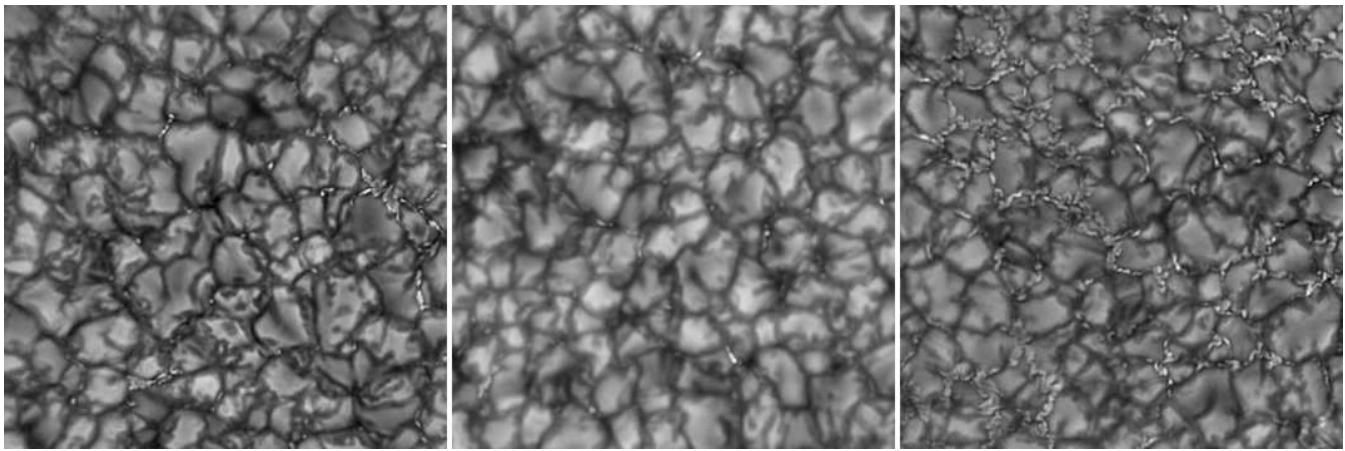}}
\caption{\sf Samples of TiO data for QS (left), CH area (middle) and AR
plage area (right). The FOV of each frame is 22''$\times$22'' or 590 x 590
pixels.}
\label{fig1}
\end{figure}

To produce one speckle reconstructed image, we need bursts of images taken in
a rapid succession. The KISIP speckle reconstruction code (Woger et al. 2008)
was applied to each burst consisting of 70 images taken with a 1~ms exposure.
All reconstructed (final) images were aligned and de-stretched. A subsonic
filter with a $v<7$ km s$^{-1}$ cut-off was applied to each final image to
remove acoustic and $f$-mode oscillations (Title et al. 1989). Typical final
data images are shown in Figure \ref{fig1}. A movie of the QS data set can be
found at the BBSO website \footnote{http://bbso.njit.edu/nst\_gallery.html}.

BPs were automatically detected in all images and then tracked from one image to
the next. Here we used the detection and tracking code previously described in
Abramenko et al. (2010). In general, our method uses the same approach as that
of Berger et al. (1998a,b), i.e., BPs are first enhanced in the images 
(by subtracting a smoothed copy of the same image) and then
selected by applying thresholding and masking.

\begin{figure}[!h]
\centerline{ \epsfxsize=7.0truein \epsffile{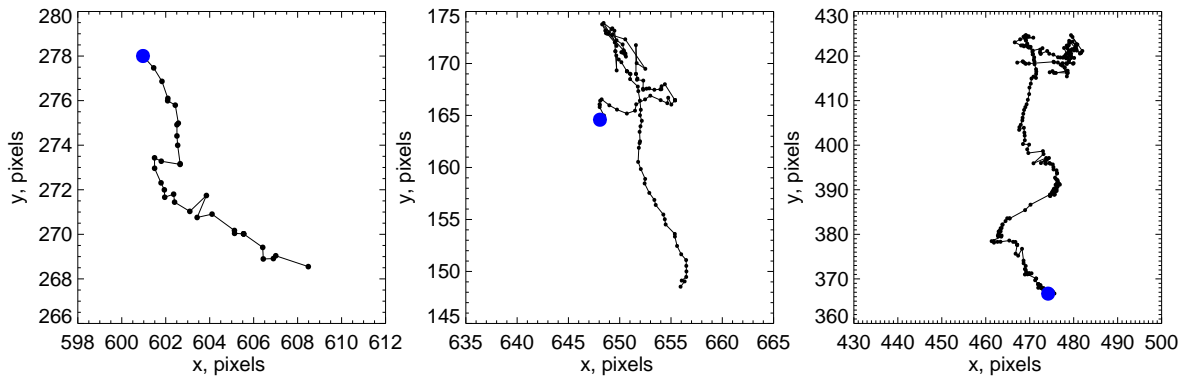}}
\caption{\sf Typical trajectories of BPs.The time intervals between adjacent
measurements (circles) is 10 seconds. Blue circles mark the start point of the
trajectory. }
\label{fig2}
\end{figure}

\begin{figure}[!h]
 \centerline{ \epsfxsize=7.0truein \epsffile{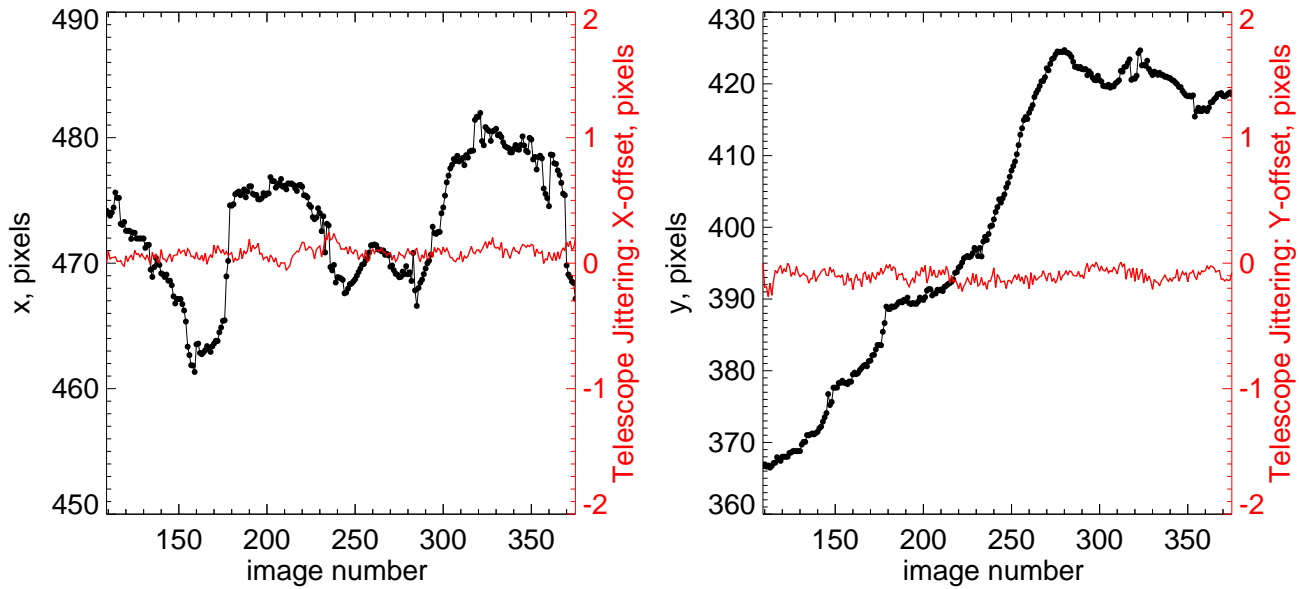}}
\caption{\sf $x$ and $y$ coordinates of the BP shown in the right panel of
Figure \ref{fig2}, overplotted with co-temporal variations of the offset between
two consecutive images (red lines, right axes). Note the scale difference
between BP's displacements (left axes)and the residual image offset (right
axes).}
\label{fig3}
\end{figure}

Typical examples of trajectories of BPs derived from the QS set are plotted in
Figure \ref{fig2}. They show that gradual drift of a BP is intermittent,
punctuated with rapid displacements. To clarify the possible reasons for these
rapid shifts, we estimated the residual image jittering in the data set by
computing an offset of each image relative to the previous image using a Fourier
aligning procedure. In Figure \ref{fig3}, we separately plot $x$ and $y$
coordinates of the trajectory shown in the right panel of Figure \ref{fig2}
along with the corresponding residual  $x$- and $y$-offsets. The figure shows
(and the calculations of the corresponding correlation coefficient confirm) that
the rapid displacements of BPs are not due to poor image alignment: image
jittering is very small and stable with the r.m.s. value of 0.1 pixel (2.7~km)
and the maximum value of 0.26 pixel. We therefore adopt an r.m.s. value of 2.7
km as a typical error of calculations of the BP position.


\begin{figure}[!h]
\centerline{ \epsfxsize=5.5truein \epsffile{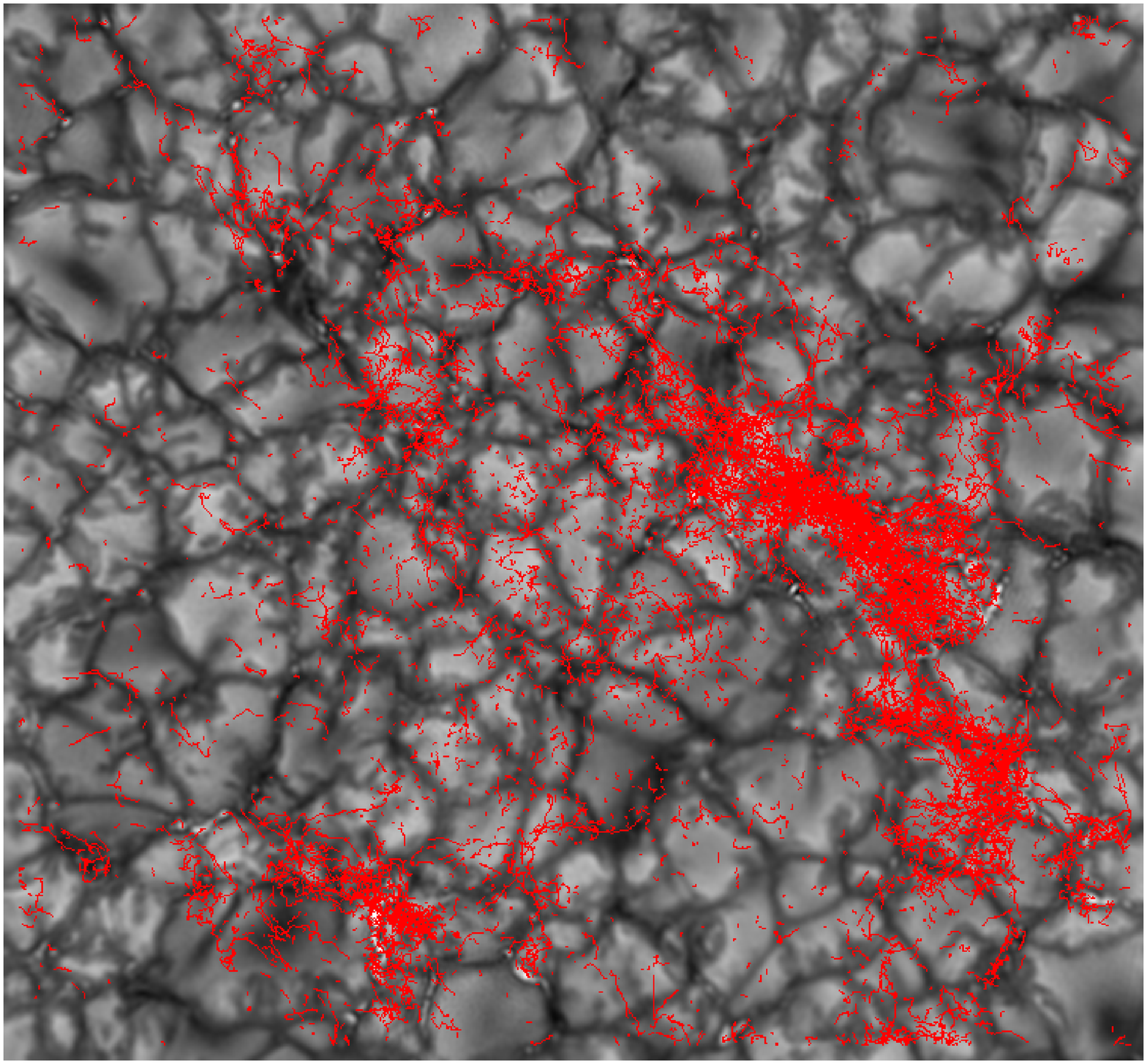}}
\caption{\sf Trajectories (red) of all BPs detected from the QS data set and
persisted longer than 3 timesteps. Background is the first image of the data
set.}
\label{fig4}
\end{figure}

In Figure \ref{fig4}, the first image in the QS data set is overplotted with all
derived BP trajectories. The BP tracks do not evenly cover the FOV (or solar
surface, for that matter) instead, they rather tend to lie along the
supergranular boundaries, i.e., along the network of magnetic fields. This
picture qualitatively agrees with previous studies (Lawrence \& Schrijver 1993;
Cadavid et al. 1999).
Namely, Lawrence \& Schrijver (1993) reported the fractal
dimension of a set of random walk sites to be $D=1.56 \pm 0.08$. In our case,
the fractal dimension of the red-colored area in Figure \ref{fig4} is slightly
lower,  $1.45 \pm 0.06$

\section{ Lagrangian Velocities }

A traditional method to describe turbulence is the Euler approach, when one
studies statistical properties of a flow by analyzing a velocity field specified
at all spatial points of the volume. This method is useful, especially in
modeling of turbulent flows. However, when we are dealing with diffusive
properties of tracers within a turbulent fluid flows, the Lagrangian approach is
more efficient then (see, e.g., Monin \& Yaglom 1975). It is based on
description of statistical properties of an individual ``fluid particle''
embedded in the turbulent flow. Such particles are assumed to be large enough,
compared to the molecular scale, and, at the same time, small enough to be
considered as solid bodies shifted as a whole (without inner deformation). The
Lagrangian approach is directly related to the real motions of the physical
particles that form the turbulent flow, and as such, it is more realistic, as
compared to the Euler method.

Photospheric BPs seem to be good candidates for such ``fluid particles'' that
are transported by turbulent photospheric flows. This approach was used, for
example, by van Ballegooijen et al. (1998) to determine the Lagrangian velocity
correlation function with subsequent application to the coronal heating problem
(Cranmer \& van Ballegooijen 2005). Here we utilize the Lagrangian method to
examine the turbulent diffusion in the quiet Sun photosphere.

Our first step was to compute the spatial displacement, $(\Delta l)_{i}$, of an
individual $i$-th BP as a function of time interval, $\tau$, measured in seconds
form the moment of the first detection of this BP. We then calculated the
average (over all BPs) displacement for each time interval to obtain the average
squared displacement as a function of time (the displacement spectrum).

\begin{equation}
(\Delta l)^{2}(\tau)=\langle|{\bf X}_{i}(0)-{\bf
X}_{i}(\tau)|^{2}\rangle,\label{Delta2}
\end{equation}
where ${\bf X}_{i}(0)=(x_{0},y_{0})$ and ${\bf X}_{i}(\tau)=(x_{\tau},y_{\tau})$
are coordinates of the $i$-th BP at the moment of its first detection and $\tau$
seconds later, respectively. The form of the function $(\Delta l)^{2}(\tau)$ is the
key for understanding the nature of the turbulent diffusion.
The power index, $\gamma$, of the displacement spectrum is defined as
\begin{equation}
(\Delta l)^{2}(\tau)\sim \tau^{\gamma}
\end{equation}
and in practice is derived as a slope of the spectra over a range of
$\tau$. This index was used to quantify the strength of diffusivity (see, e.g.,
Cadavid et al. 1999). 

A broad definition of $\gamma$ for various statistical
moments was discussed and applied for solar data by Lawrence et al. (2001). Here
we utilize only the second statistical moment of displacements.

\section{ Diffusion Regime in the Photosphere as Derived from Various Sources}

\begin{figure}[!h]
\centerline{ \epsfxsize=6.0truein \epsffile{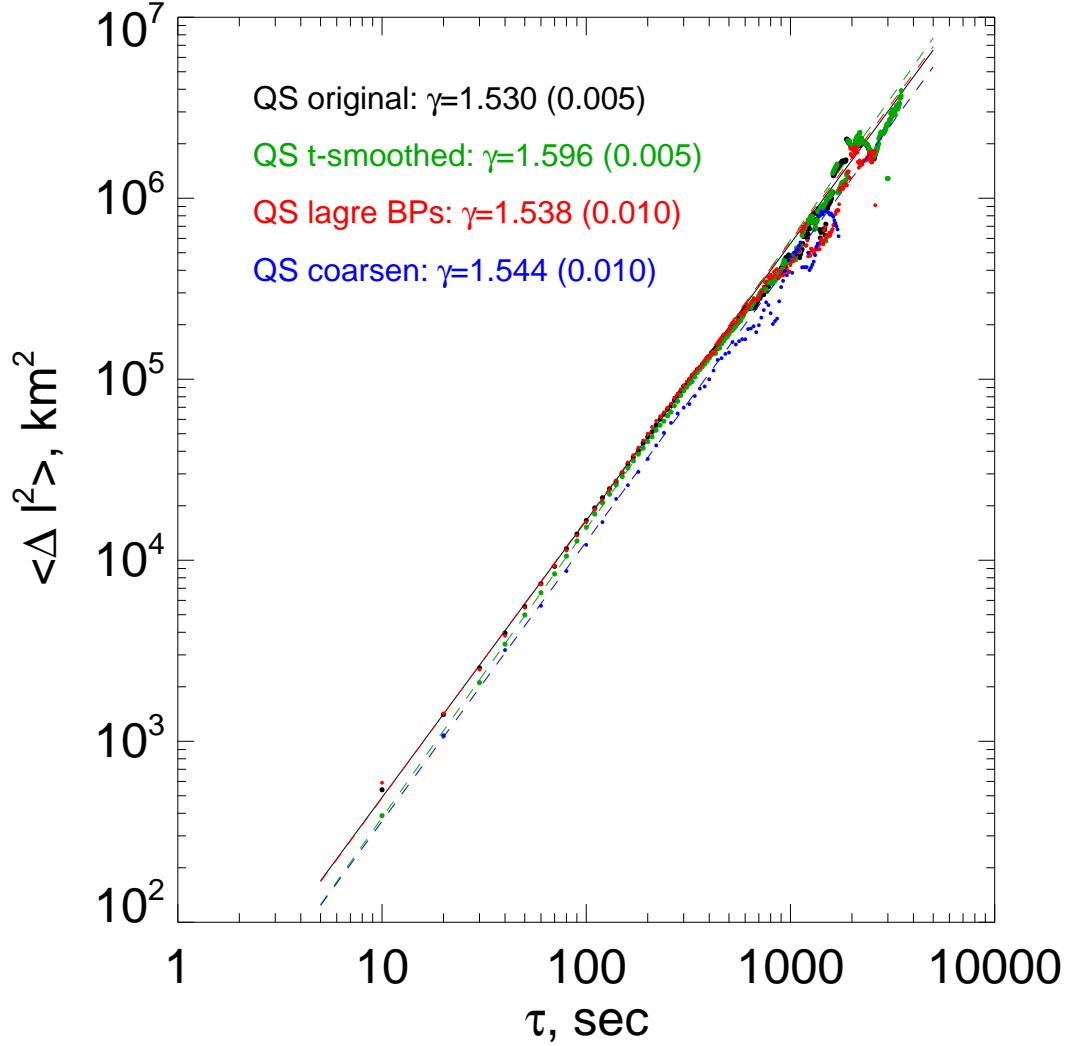}}
\caption{\sf Displacements $(\Delta l)^{2}(\tau)$ computed according to Eq.
\ref{Delta2}. Black data symbols represent results derived from the original QS
data set,  green colors represent results derived from the time-smoothed data
set, red colors are plot results derived from a subset of only large, bright and
long-live BPs, while blue symbols represent  the result for the most degraded
version of the data set(2$\times$2 re-binned and Gaussian-deconvolved data set
of 20~s cadence). In all cases, the error bars did not exceed the size of the
plotting symbols. In all experiments, slope $\gamma$ was calculated as the best
linear fit inside the range of 10-300~s. One-$\sigma$ errors of the fitting are
shown in brackets.}
\label{fig5}
\end{figure}

To understand how our results could be affected by the resolution of the data
and its quality, we performed the following tests. First, using the QS data set,
we calculated $(\Delta l)^{2}(\tau)$ in four different ways. Results are shown
in Figure \ref{fig5}. There, the black dots and black solid straight line
represent results for 7912 BPs derived from the final QS data set as described
in the Data section: i.e., speckle reconstructed, aligned, de-stretched,
subsonic-filtered images with 10 seconds time cadence utilized to detect BPs
with an area of no less than 3 squared pixels, that lasted more than 3 time
steps. The power law index, $\gamma$, calculated as the best linear fit to the
data points within the 10-300~s  time interval, was found to be 1.530$\pm$0.005.

Next, we degraded the data set by applying a 3-point running-box averaging along
the time axis (green circles and lines  in Figure \ref{fig5}), thus essentially
eliminating the inhomogeneity of the intensity distribution (important to
determine the location and the center of a BP). We thus detected total 6661
events and the power law index in this case increased only slightly relative to
the un-degraded original data (from 1.53 to 1.60).

We repeated the same calculations taking into account only large, bright and
long living BPs (red color symbols and line in Figure \ref{fig5}). The BP area
threshold was set here to 8 pixels, the life time threshold to 10 time steps
(100~s), and the brightness threshold to 90 DN (instead of 85 DN). Under these
requirements, we detected only 1861 tracked BPs. In this test run, all possible
uncertainties arising from analyzing low intensity, small, short living BPs
(which actually comprise the majority of the detected BPs) were eliminated. The
results, however, were not much different: the power index, $\gamma$, determined
within the same time interval of 10-300~s was  $1.538\pm0.010$. At larger time
scales (above approximately 700 s), the difference with the previous experiments
is more noticeable, but still negligible.

In our final experiment (blue color data points in Figure \ref{fig5}), we first
re-binned the original QS data set to the 2$\times$2 original pixels mesh, and
then convolved it with a Gaussian function to mimic 0.$''$4 spatial resolution.
Afterward, we discarded every even image in the data set to increase the time
cadence to 20~s. In doing so, we degraded the resolution and time cadence of our
data set to closely match that utilized in Cadavid et al. (1999). The intensity
threshold for BP detection was set to 105 DN, so that only brightest BPs were
taken into account. The total number of detected tracked BPs was only 840. The
blue line in Figure \ref{fig5} that corresponds to this experiment, is the
lowest one, which means that on average the displacements are smaller.  However,
the power index $\gamma=1.544\pm0.010$, which is very similar to the value of
$\gamma=1.530\pm0.005$ inferred from the un-degraded original data set.

Subsonic filtering did not significantly affect our results, we obtained
$\gamma=1.535\pm0.008$ for the non-filtered QS data set. 

The above exercise shows that index $\gamma$ determined for small time scales
($<$ 300~s) does not seem to depend on data processing and apparently reflects
an intrinsic property of BP's motions recorded in the data set: a
super-diffusion turbulent regime with $\gamma\approx1.5$. Only on larger time
scales ($>$ 600~s), the function $(\Delta l)^{2}(\tau)$ depends slightly on data
processing and/or spatial resolution.

\begin{figure}[!h]
\centerline{\epsfxsize=6.0truein \epsffile{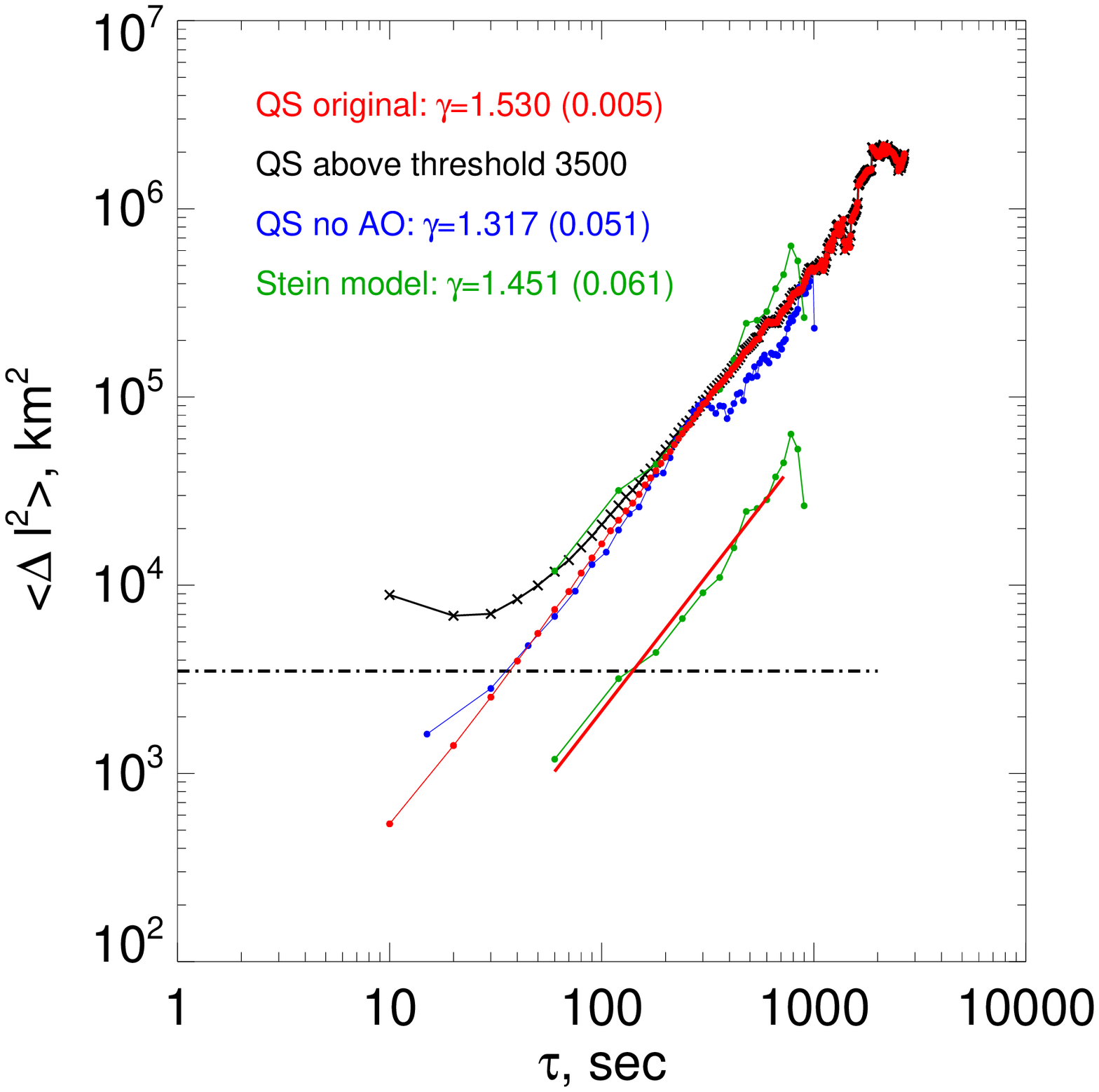}}
\caption{\sf Function $(\Delta l)^{2}(\tau)$ derived from i) the original QS
data set (red), ii) small displacements (below a 3500 km$^{2}$ threshold, black
dash-dotted line) in the original QS data set were discarded (crosses); 
 iii) July 29,
2009 NST data set recorded without an AO correction (blue), and iv) model solar
granulation data (green, Stein et al. 2007). For better viewing, the model data are
 also shifted down and their best linear fit is shown with the thick red straight line. }
\label{fig8}
\end{figure}

Another reason for the discrepancies may be that the NST data enable us to
measure much smaller BP displacements than the SVST observations do. To explore
this possibility, we discarded all squared displacements $(\Delta l)^{2}(\tau)$
smaller than the SVST squared displacement detection threshold, 3500 km$^{2}$
(see Fig. 1 in Cadavid et al. 1999). The result is shown in Figure \ref{fig8}
(crosses). At small scales ($\tau<200$~s), data points are well above the
original data points (red). 
Interestingly, for the time scales of 40-300~s, the slope of the spectrum
is $1.26 \pm 0.02$, which is very close to that reported by Lawrence et al.
(2001) from the SVST data.

%

Two additional tests described below further confirm validity of
super-diffusion regime and the relevance of data quality for measuring the
index, $\gamma$, of the displacement spectra.

\begin{figure}[!h]
\centerline{\epsfxsize=7.0truein \epsffile{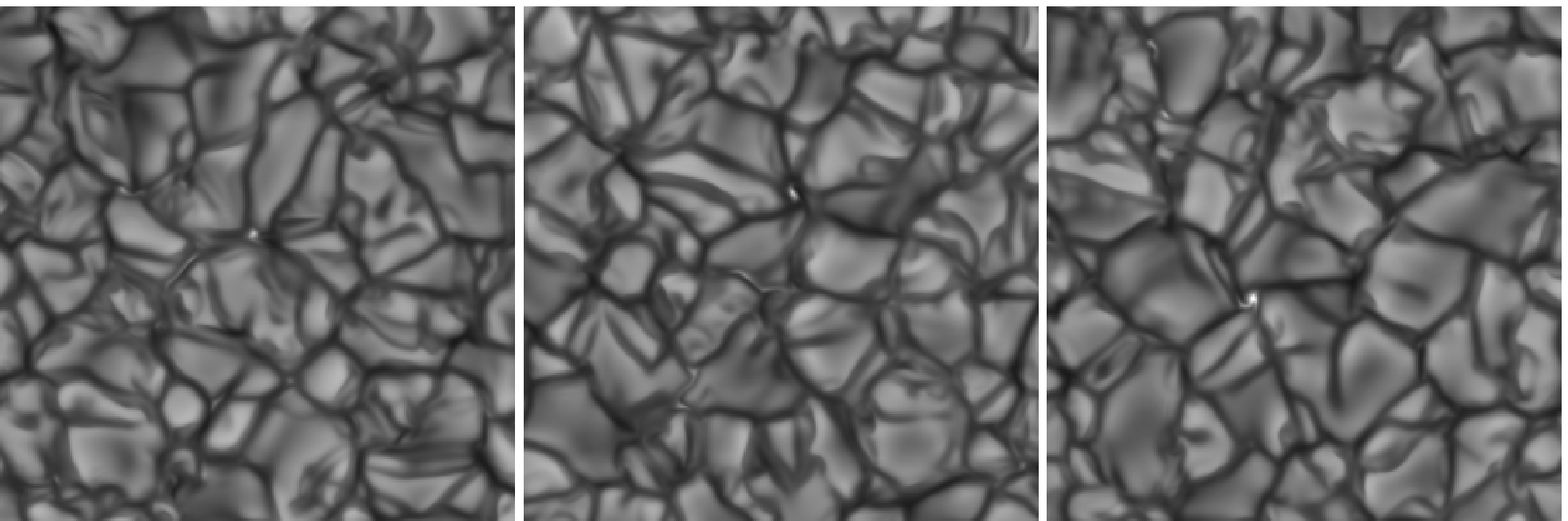}}
\caption{\sf Three fragments of simulated solar granulation images obtained from a 3D MHD solar
magneto-convection code (Stein et al. 2007). BPs in intergranule lanes are visible.}
\label{fig10}
\end{figure}

First, we applied our code to model data of solar granulation obtained from
3D MHD solar magneto-convection simulations (Stein et al. 2007). The model data
set consisted of 146 intensity images of 1008$\times$1008 pixels (48$\times$48
Mm) separated by a 60~s time intervals. Three fragments of the simulated
intensity images are shown in Figure \ref{fig10}. The surface density of BPs in
this case was lower than that for the observed QS data. The model data show
super-diffusion with $\gamma=1.451\pm0.061$ inside the time interval of 60 -
660~s (Figure \ref{fig8}, green symbols), which is in good agreement with
observations. 

In the second run, we applied our code to an NST data set obtained on July 29,
2009 (see Goode et al. 2010b for description of this data set). The main reason
to use this set is that it is of inferior quality as compared to the August 3,
2010 data utilized in this study. In particular, the July 29, 2009 data were
recorded under good seeing conditions without the AO correction, and the time
coverage was four times shorter. The displacement spectrum obtained from these
inferior data (Figure \ref{fig8}, blue symbols) has poorer linearity, less
pronounced super-diffusion regime and narrower range of super-diffusion. The
spectral index $\gamma$ is equal to $1.317\pm0.051$ inside the time interval of
15 - 210~s.

All the tests performed in this section point to a conclusion that
spatial and temporal resolution affect the displacement spectrum and its slope.
Higher precision measuring positions of BPs in the NST data appears to be the
main reason for the persistent detection of super-diffusion regime in the
photosphere.

\section{Turbulent Diffusion in Quiet Sun, Coronal Hole and AR Plage Areas}

\begin{figure}[!h]
\centerline{ \epsfxsize=6.0truein \epsffile{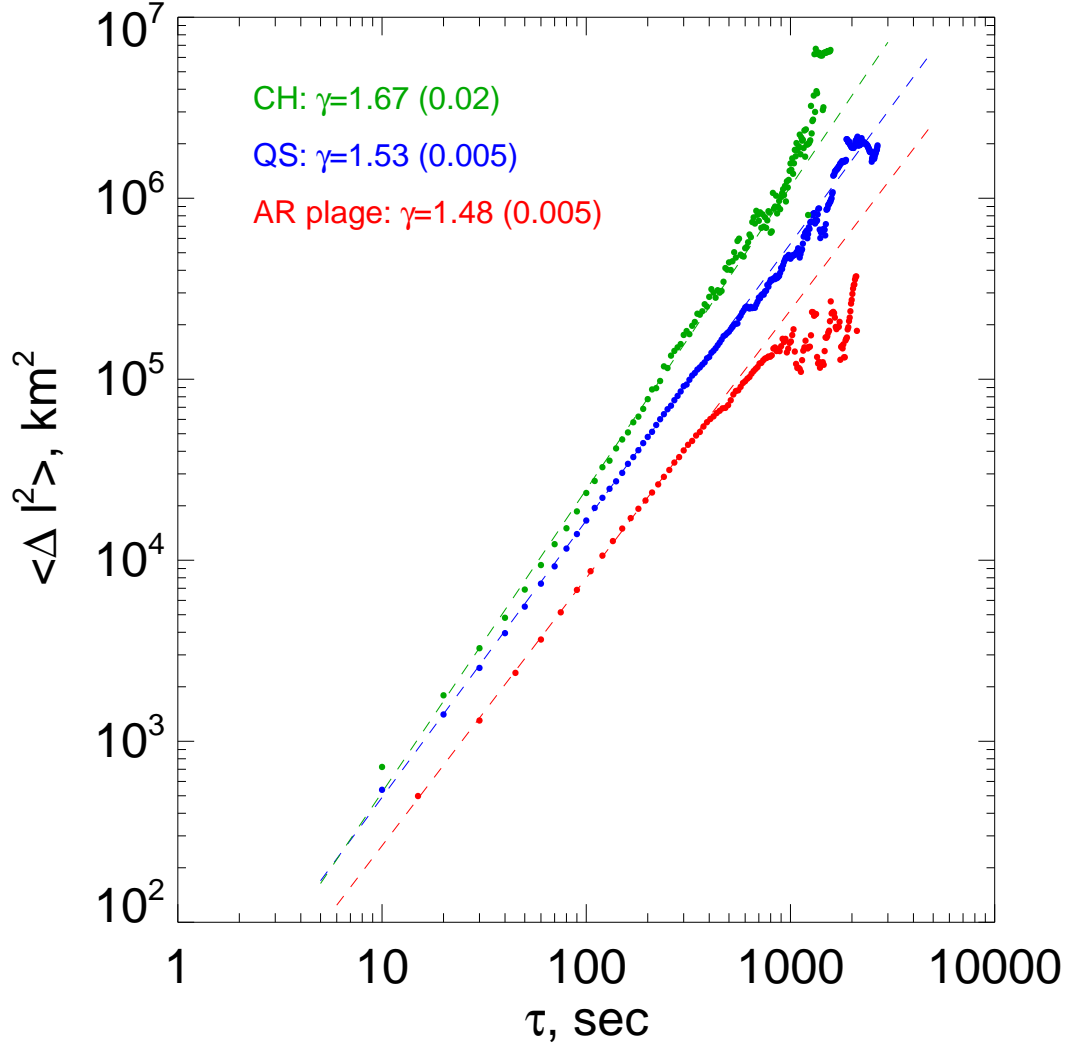}}
\caption{\sf Function $(\Delta l)^{2}(\tau)$ determined for the CH data (green),
QS area (blue) and AR plage area (red), according to Eq.\ref{Delta2}. The error
bars do not exceed the size of plotting symbols. Other notations are the same as
in Figure \ref{fig5}.}
\label{fig6}
\end{figure}

\begin{figure}[!h]
\centerline{ \epsfxsize=6.0truein \epsffile{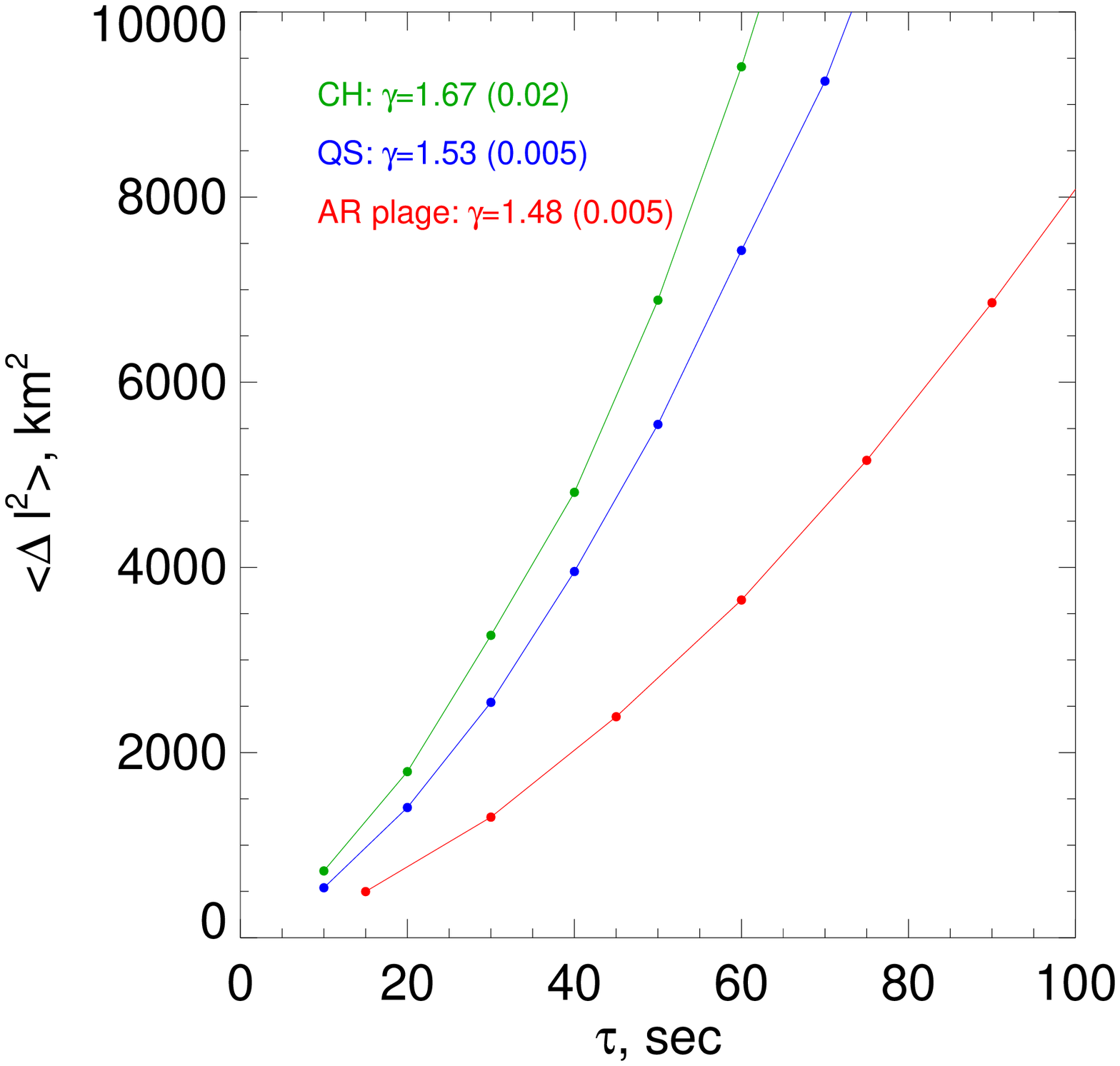}}
\caption{\sf The short time scale end of the plot in Figure \ref{fig6} plotted
in linear axes. Notations are the same as in Figure \ref{fig6}.}
\label{fig7}
\end{figure}

In this section, we discuss the displacement spectra for the CH and ARP data
and compare them with the displacement spectrum for the QS data set (Figure
\ref{fig6}). Although the three spectra display noticeable difference, they
demonstrate that the super-diffusion regime ($\gamma>1$) is present in all
three magnetic areas. In Figure \ref{fig7}, we plot the same spectra in linear
coordinates for a short time range only.
%

%
The ARP spectrum shows the shallowest slope and smallest displacements,
indicating the lowest, among these three samples, level of turbulent diffusion.
The CH data show the steepest spectrum and largest displacements. This was
expected since BPs in the AR plage area are the most crowded and tightly packed
in the narrow inter-granular lanes, when compared to the QS and CH environment
(see Figure \ref{fig1}). In the CH, BPs have the most freedom to move, because
they have a lower population density. Indeed, from a single CH image, we find
the BP surface density to be between 0.08 and 0.65 BP per Mm$^{2}$, while this
density is 0.28-0.72 Mm$^{-2}$ and 0.71-1.78 Mm$^{-2}$ in the QS and the AR
plage area, respectively. These density values are comparable with those
reported in other studies (e.g., Sanchez Almeida et al. 2010). Therefore,
turbulent diffusion is most enhanced in a coronal hole on all time scales.

\section{Turbulent Diffusion Coefficient as a Function of Scale  }

Monin \& Yaglom (1975) argued that in case of fully developed turbulence,
the coefficient of turbulent diffusion of a passive scalar should depend on 
temporal and spatial scales. In 2D turbulence (in our case - 
solar surface), the expression for the turbulent diffusion coefficient is 
(see Monin \& Yaglom 1975):
\begin{equation}
K(\tau)=\frac{1}{4}\frac{d}{d\tau}\langle l^{2}(\tau)\rangle,
\label{Monin_K}
\end{equation}
where $\langle l^{2}(\tau)\rangle$ is a variance of  ``fluid particle''
displacements for given $\tau = t-t_{o}$. Note that $\langle
l^{2}(\tau)\rangle$ is equivalent to function $(\Delta l)^{2}(\tau)$, discussed
above, which can be approximated, at a given range of scales (Sections 4 and 5),
by
\begin{equation}
(\Delta l)^{2}(\tau)=c\tau^{\gamma},
\label{gen}
\end{equation}
where $c=10^{y_{sect}}$ and $\gamma$ and $y_{sect}$ can be derived from the
best linear lit to the data points plotted in a double-logarithmic plot. Then
the diffusion coefficient can be written as
\begin{equation}
K(\tau)=\frac{c\gamma}{4}\tau^{\gamma-1}.
\label{Ktau}
\end{equation}
When $\tau$ is excluded from Eqs. \ref{Monin_K} and \ref{gen}, we obtain a
relationship between the diffusion coefficient and a spatial scale:
\begin{equation}
K(\Delta l)=\frac{c\gamma}{4}((\Delta l)^{2}/c)^{(\gamma-1)/\gamma}.
\label{K_l}
\end{equation}

\begin{figure}[!h]
\centerline{
\epsfxsize=3.5truein\epsffile{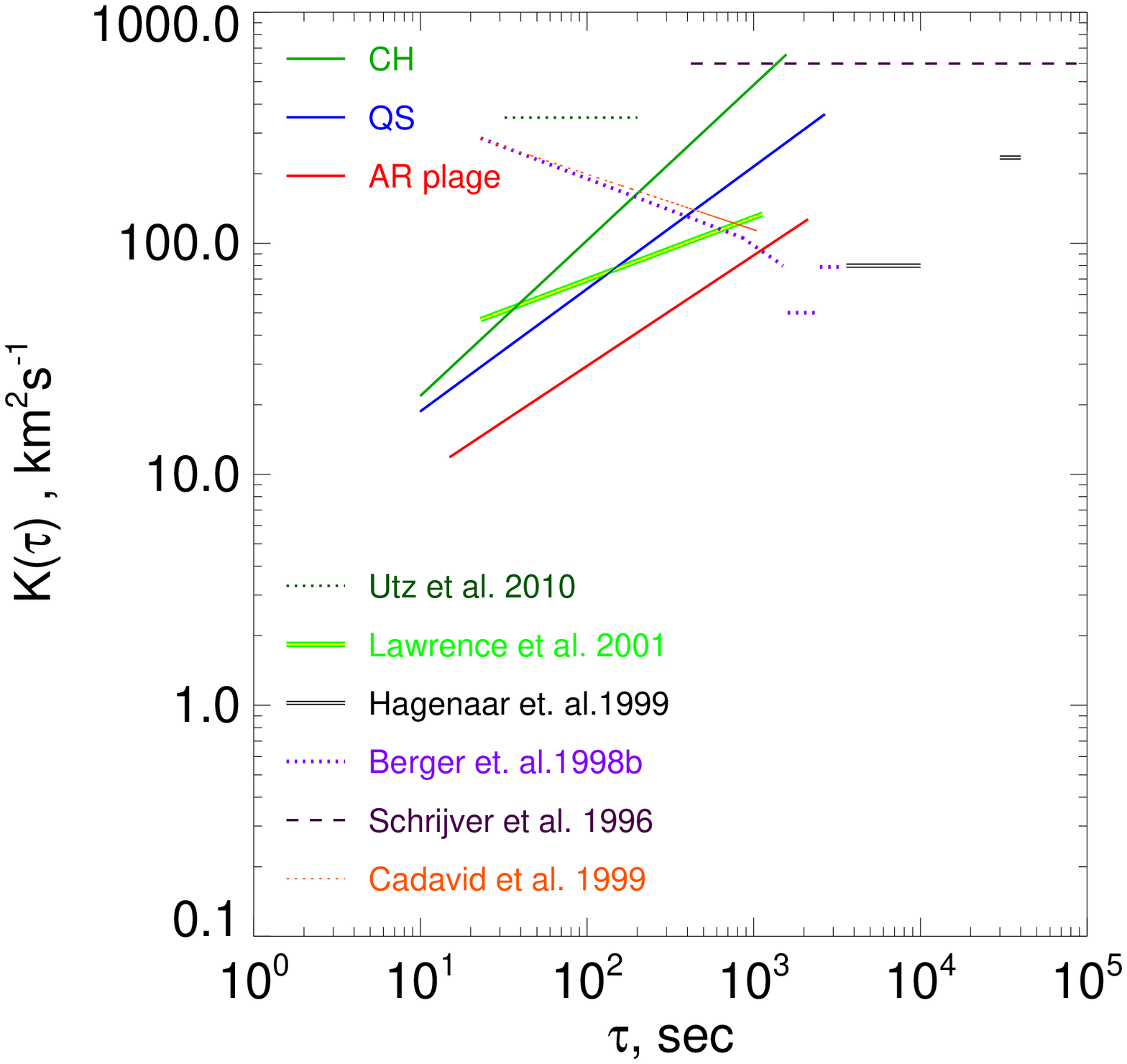}
\epsfxsize=3.5truein\epsffile{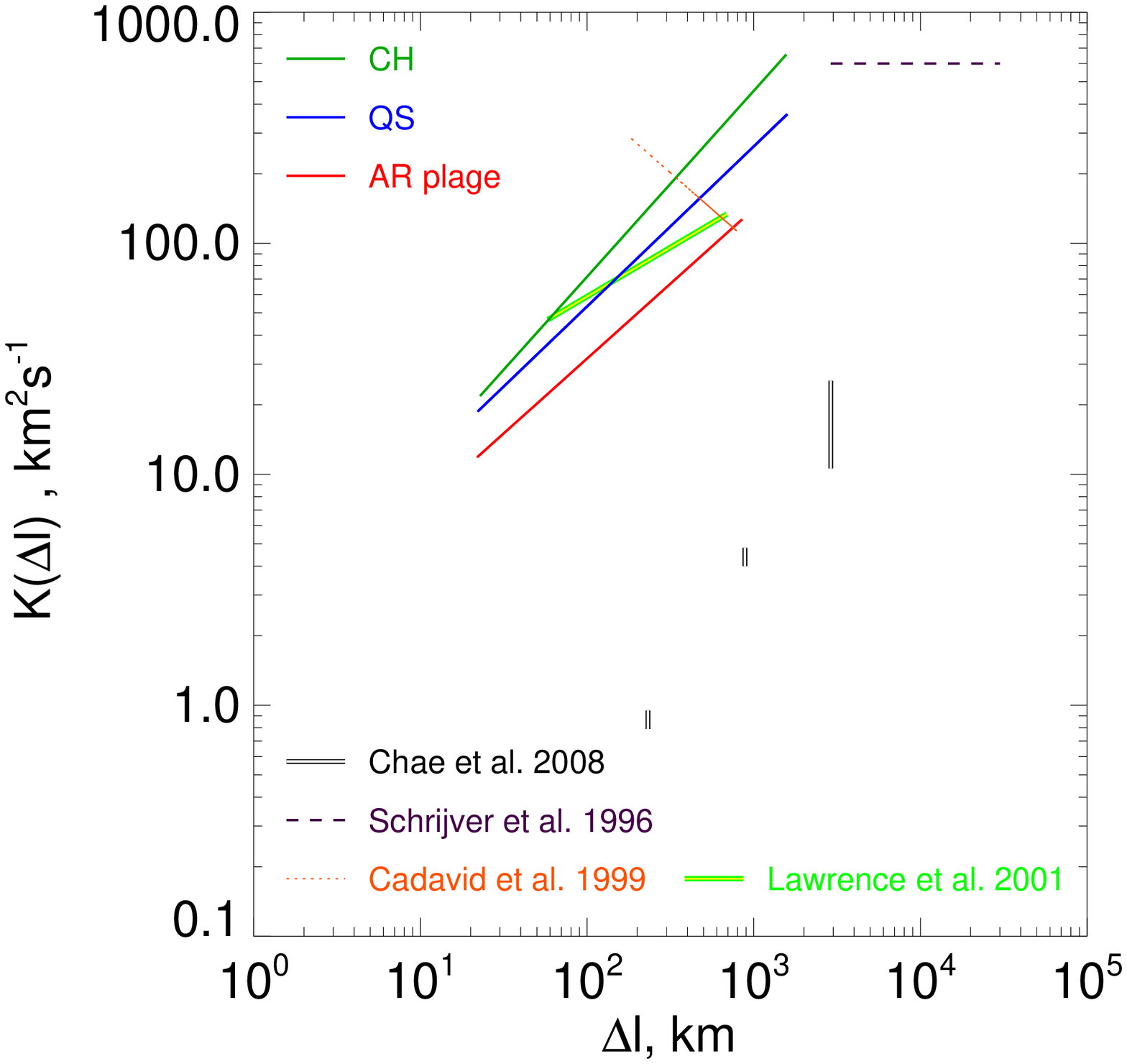}}
\caption{\sf Turbulent diffusion coefficient plotted as a function of temporal
(left) and spatial (right) scales. Solid lines refer to the NST data shown in
Figures \ref{fig6} and \ref{fig7} with the same color notations.}
\label{fig9}
\end{figure}

Figure \ref{fig9} (solid lines) plots $K(\tau)$ and $K(\Delta l)$ calculated
from the three linear fits presented in Figure \ref{fig6} by means of Eqs.
\ref{Ktau} and \ref{K_l}. The largest diffusion coefficients are measured in the
CH, while the smallest ones are detected in the AR plage area, which is in
accordance with the strength of the super-diffusion regime in these areas
(magnitude of $\gamma$) . In general, we find that as the temporal and spatial
scales decrease, the diffusion coefficient decreases, too. This holds true only
in the case of super-diffusion: indeed, as it follows from Eqs. \ref{Ktau} and
\ref{K_l}, for all $\gamma>1$ both $K(\tau)$ and $K(\Delta l)$ are monotonically
increasing functions. When $\gamma=1$, the diffusion coefficient is constant at
all scales (Brownian motion, or random walk), while in the case of sub-diffusion
($\gamma<1$), the diffusion coefficients decrease with increasing scales.

\section{Summary and Discussion}

We report the presence of the super-diffusion regime in the displacement of
photospheric BPs measured in a quiet-sun area, a coronal hole, and in an active
region plage. Super-diffusion, measured by the index $\gamma$ of the squared
displacement spectrum, is weakest in the plage area ($\gamma=1.48$) and
strongest in the coronal hole area ($\gamma=1.67$). The data quality (spatial
resolution, time cadence, alignment) has a direct effect on the result: the
better the data quality the more pronounced the super-diffusion regime is. The
super-diffusion regime was also detected in the 3D MHD model data of the solar
magneto-convection performed by Stein and co-authors (Stein et al. 2007).

We obtained an analytical expression for the turbulent diffusion coefficient as
a function of scales and $\gamma$. For the case of super-diffusion, the
coefficient of turbulent diffusion is directly proportional to both temporal and
spatial scales. At the minimal spatial (22 km) and temporal (10 sec) scales
considered here the diffusion coefficient in the CH and QS areas was found to be
22 and 19~km$^{2}$~s$^{-1}$, respectively. Whereas in the AR plage the
coefficient was about 12 km$^{2}$ s$^{-1}$ for the minimal temporal scale of
15~s.

We compare our findings to previous studies of the photospheric diffusion
in the quiet sun (see Figure \ref{fig9}).

Hagenaar et al. (1999) tracked magnetic elements using SOHO/MDI high resolution
magnetograms and reported a diffusion coefficient increasing with scale 
(double lines in the left panel of Figure \ref{fig9}). 

Chae et al. (2008) determined the diffusion coefficient from observed dynamics
of photospheric magnetic fields by solving the equation of magnetic induction.
These authors analyzed plage and network areas by utilizing SOHO/MDI full disk and
high resolution magnetograms as well as Hinode/SOT data. They reported the
smallest magnitudes of $K$, to date, for three spatial scales (double lines in the
right panel of Figure  \ref{fig9}). It is worth noting that these three data
points are in a very good agreement with both Kolmogorov and
Iroshnikov-Kraichnan scaling (see Figure 4 in Chae et al. 2008).

Utz et al. (2010) estimated the diffusion coefficient in the framework of the
normal diffusion paradigm, and therefore, their results do not depend on scale
(green dots in the left panel of Figure  \ref{fig9}). They reported the value of
$K$ = 350$\pm 20$ km$^{2}$ s$^{-1}$ for the time interval of 32 - 200 s.

Schrijver et al. (1996) reviewed earlier data on the diffusion coefficient and
argued that the reported values, which are around 200 km$^{2}$ s$^{-1}$, should
be multiplied by a factor of 2-3 to account for the high mobility of small
(unresolved) magnetic features that may contribute significantly in the
resulting diffusivity. The corrected diffusivity of about 600 km$^{2}$ s$^{-1}$
(dashed lines in Figure \ref{fig9}) is then in a good agreement with the
well-performing model for the magnetic flux transport during a solar cycle
(Wang \& Sheeley 1994).

Berger at al. (1998b) analyzed a G-band data set obtained with the SVST on
October 5, 1995 (the same data set was utilized later in Cadavid at al. 1999 and
Lawrence et al. 2001). The authors used a BP tracking method (L{\"o}fdahl et al.
1998, Berger et al. 1998a). They found a nearly constant diffusion coefficient
of 50 km$^{2}$ s$^{-1}$ inside the 1600-2400~s time interval and 79 km$^{2}$
s$^{-1}$ inside the 2500-3400~s interval (blue dots in the left panel of Figure
\ref{fig9}). For shorter ($<1500$~s) time scales, the authors reported a
diffusion coefficient decreasing with spatial scale, and they thus concluded
that a normal diffusion approximation is not applicable for short time scales.

We also used the displacements spectra shown in Figure 1 of Cadavid et
al. (1999) and in Figure 2 of Lawrence et al. (2001) to calculate the diffusion
coefficient by means of Eqs. \ref{Ktau} and \ref{K_l} (the orange dots and the
green double line, respectively, in Figure \ref{fig9}). Cadavid et al. (1999)
displacements spectrum produced a diffusion coefficient that is inversely
proportional to scales (both temporal and spatial), indicating a dominant
sub-diffusion regime. Lawrence et al. (2001) displacements spectrum produced a
diffusion coefficient that increases with scales, however the rate of increase
is slower than that derived from the NST data. 

The turbulent diffusion coefficients were derived in this study from the second
moment of displacements (see Eqs. \ref{Ktau} and \ref{K_l}). Lawrence et al.
(2001) utilized a general expression for the diffusivity index, $\gamma(q)$,
treating it as a function of the statistical moment, $q$ : \begin{equation}
\langle (\Delta l)^{q} \rangle ^{1/q} \sim \tau^{\gamma(q)/2}. \label{Lawr}
\end{equation} These authors found that in the limit of a very large data set,
the magnitude of $\gamma$ tends to be independent of $q$. Moreover, they showed
that errors in calculating of $\gamma$ become significantly smaller when $q \to
0$. Therefore, it is useful to derive $\gamma$ for small $q$ and compare it with
that obtained for $q=2$. From our QS data set, we derived $\gamma (0.1)= 1.580
\pm 0.0035$, which is only 3\% higher than $\gamma (2)= 1.530 \pm 0.005$. Note
that from the SVST data, Lawrence et al. (2001) obtained $\gamma (0.1)= 1.27 \pm
0.01$ and $\gamma (2)= 1.13 \pm 0.01$, and they differ from each other by 12\%.
Moreover, the obtained here magnitude of $\gamma$ is in a very good agreement
with that predicted from the CTRW approach, $\gamma=1.54$ (Lawrence et al.
2001).

Modern models of the small-scale turbulent dynamo in the photosphere (Boldyrev
\& Cattaneo 2004; V{\"o}gler \& Sch{\"u}ssler 2007; Pietarila Graham et al.
2009) utilize the collisional value of magnetic diffusivity (0.01 - 10
km$^2$s$^{-1}$) based on the electric conductivity in the photosphere. At the
same time, utilizing the turbulent magnetic diffusivity would be more justified
physically, as long as the turbulent diffusivity determines the minimum scale
for magnetic elements. The measured so far value of turbulent magnetic
diffusivity (70-350 km$^2$s$^{-1}$), being interpreted as a scale-independent
parameter, leaves a very slim chance to successfully model the small-scale
turbulent dynamo in the photosphere. Thus, in the case of very high diffusivity
on very small scales (sub-diffusivity),  chances for tiny magnetic field
concentrations to resist the spreading action of turbulent flows are small, so
that the dynamo is restrained. A super-diffusion regime on very small scales is
very favorable for pictures assuming the turbulent dynamo action since it
assumes decreasing diffusivity with decreasing scales. The idea of a small-scale
turbulent dynamo operating in the quiet sun photosphere received strong
observational support when turbulent magnetic fields were discovered with with
Hinode instruments (Centeno et al. 2007; Lites et al. 2008; Orozco Suarez et al.
2008; Sch{\"u}ssler \& V{\"o}gler 2008).

Author are thankful to BBSO instrument team and observers for their contribution
to data acquisition. We are obliged to the anonymous referee whose comments
helped much in improving the manuscript. VA work was partially supported by NSF
grant ATM-0716512. VY acknowledges support from NASAs GI NNX08AJ20G and LWS
TR\&T NNG0-5GN34G grants. PG, VA and VY are partially supported by NSF
(AGS-0745744), NASA (NNY 08BA22G) grants. PG is partially supported by AFOSR
(FA9550-09-1-0655).  RFS was partially supported by NASA grants NNX07AH79G and
NNX08AH44G and NSF grant AST0605738.  The simulations were performed on the
Pleiades supercomputer of the NASA Advanced Supercomputing Division.

\end{document}